\documentclass[showpacs,preprintnumbers,amsmath,amssymb]{revtex4}
\usepackage{chemarr}
\usepackage{graphicx}
\usepackage{bm}  
\usepackage{amssymb}
\usepackage{color}
\usepackage{amscd}

\begin{document}

\title[Stochastic Low Reynolds Number Swimmers]{Stochastic Low Reynolds Number Swimmers}

\author{Ramin Golestanian}

\address{Department of Physics and Astronomy, University of Sheffield, Sheffield S3 7RH,
UK} \email{r.golestanian@sheffield.ac.uk}

\author{Armand Ajdari}
\address{Gulliver, UMR CNRS 7083, ESPCI, 10 rue Vauquelin, 75005
Paris, France}

\begin{abstract}
As technological advances allow us to fabricate smaller autonomous
self-propelled devices, it is clear that at some point directed
propulsion could not come from pre-specified deterministic periodic
deformation of the swimmer's body and we need to develop strategies
to extract a net directed motion from a series of random transitions
in the conformation space of the swimmer. We present a theoretical
formulation to describe the ``stochastic motor'' that drives the
motion of low Reynolds number swimmers based on this concept, and
use it to study the propulsion of a simple low Reynolds number
swimmer, namely, the three-sphere swimmer model. When the
detailed-balanced is broken and the motor is driven out of
equilibrium, it can propel the swimmer in the required direction.
The formulation can be used to study optimal design strategies for
molecular-scale low Reynolds number swimmers.

\end{abstract}

\vspace{2pc}

\maketitle

\section{Introduction}

Biological molecular motors \cite{howard} are ingenious nano-scale
machines that convert chemical energy into directed mechanical work
amid strong thermal fluctuations. With the current miniaturization
trend in technology, one naturally wonders if it is possible to
synthesize devices with similar functionalities \cite{leigh-etal}.
In particular, it is desirable as a first step to design autonomous
small scale swimmers, which could later on be steered by coupling to
a guiding network or system. These swimmers could be used in
carrying cargoes or stirring up fluids at small scales.

There is a significant complication in designing swimmers at small
scale as they have to undergo non-reciprocal deformations to break
the time-reversal symmetry and achieve propulsion at low Reynolds
number \cite{taylor}. While it is not so difficult to imagine
constructing motion cycles with the desired property when we have a
large number of degrees of freedom at hand---like nature does, for
example---this will prove nontrivial when we want to design
something with only a few degrees of freedom and strike a balance
between simplicity and functionality, like most human-engineered
devices \cite{purcell1}. Recently, there has been an increased
interest in such designs
\cite{3SS,josi,drey,igor,lee,ali2,feld,holger,lesh,peko,yeomans1,anna,yeomans2,yeomans3,yeomans4,lauga2,ag-pre,epje,ag}
and two interesting examples of such robotic micro-swimmers have
been realized experimentally using magnetic colloids attached by
DNA-linkers \cite{Dreyfus,Pietro}. Among others, a simple swimmer
model based on spheres connected by arms that do not interact with
the fluid \cite{3SS} has been recently used for a number of studies
including scattering of two swimmers \cite{yeomans2,yeomans4},
collective hydrodynamic coupling of swimmers \cite{yeomans3,lauga2},
general feasibility of various design properties of swimmers
\cite{ag-pre}, and the effect of large cargos on the performance of
swimmers \cite{epje}. While constructing small swimmers that
generate surface distortions is a natural choice, it is also
possible to take advantage of the general class of phoretic
phenomena to achieve locomotion---as they become predominant at
small scales---as recent experimental \cite{paxton,ozin,mano,jon}
and theoretical \cite{gla-1,gla-2,kapral} works have demonstrated.

Here we construct a general statistical mechanical formulation for
studying low Reynolds number swimmers that undergo conformational
changes in a stochastic manner pertinent to systems of molecular
scale. We attribute transition rates to each deformation move or
swimming stroke, and calculate the propulsion velocity as a function
of these rates. Our formulation provides a general prescription on
how to construct the relevant portions of the configurational space
of swimmers, and how to take advantage of the complexities in this
space to maximize the efficiency of the swimmer. We apply the
formulation to the specific example of the three-sphere swimmer
model, which yields interesting results.

The rest of the paper is organized as follows: Section
\ref{sec:hydro} describes the general formulation of hydrodynamics
of low Reynolds number swimmers, and it is followed by Section
\ref{sec:kinetics} that is devoted to the statistical mechanics of
the conformational changes in swimmers. The formulation is applied
to the example of three-sphere swimmer model in Section
\ref{sec:3SS}, which is followed by concluding remarks in Section
\ref{sec:conc}.

\section{Hydrodynamics of Low Reynolds Number Swimming}
\label{sec:hydro}

We consider a deformable extended body as a system composed of $N$
point-like solid components described by their position vectors
${\bf r}^{\alpha}(t)$. The deformation is related to internal forces
exerted between these solid components, so that on component
${\alpha}$ is exerted a net force ${\bf f}^{\alpha}(t)$, that is in
turn applied oton the fluid at ${\bf r}^{\alpha}(t)$. In our
description of point-like objects, the hydrodynamic interactions
between the objects relate these force to the velocities of the
components ${\bf v}^{\alpha}(t)={\dot {\bf r}}^{\alpha}(t)$ via the
Oseen tensor ${\cal H}_{ij}({\bf r},{\bf r}')$ \cite{Oseen} (roman
indices describe spatial components), namely
\begin{equation}
v^{\alpha}_{i}=\sum_{\beta} M^{\alpha \beta}_{ij}
f^{\beta}_{j},\label{eq:Oseen-1}
\end{equation}
where $M^{\alpha \beta}_{ij}={\cal H}_{ij}({\bf r}^{\alpha},{\bf
r}^{\beta})$ and summation over repeated roman indices that define
the vector components is understood. The Oseen tensor is the Green
function for the Stokes equation with the appropriate boundary
conditions and its explicit form depends on the problem we are
considering. For example, in the simplest case we can treat the
solid particles as point-like and use the $1/r$-type expressions for
the off-diagonal components of the Oseen tensor, while putting in
$1/(6 \pi \eta a)$ for the diagonal components where $a$ is the
radius of the particles and $\eta$ is the viscosity of water. If
necessary, one could also incorporate finite size corrections and
the effect of confining boundaries by using the appropriate form of
the Green function.

We can now invert equation (\ref{eq:Oseen-1}) as
\begin{equation}
f^{\alpha}_{i}=\sum_{\beta} N^{\alpha \beta}_{ij}
v^{\beta}_{j},\label{eq:f=}
\end{equation}
where $N^{\;\alpha \beta}_{ij}$ is the resistance (friction) tensor
that satisfies $\sum_{\beta} M^{\alpha \beta}_{ij} N^{\beta
\gamma}_{jk} =\delta_{\alpha \gamma} \delta_{i k}$.

For a swimmer that is not subjected to external forces, the local
and instantaneous forces in the body are subject to the constraint
\begin{equation}
\sum_{\alpha} {\bf f}^{\alpha}=0,\label{eq:force-free}
\end{equation}
which yields
\begin{equation} \sum_{\alpha,\beta} N^{\alpha
\beta}_{ij} v^{\beta}_{j}=0.\label{eq:oseen-2}
\end{equation}
Similarly, if the swimmer is not under the effect of a net external
torque, an additional constraint applies
\begin{equation}
\sum_{\alpha} ({\bf r}^{\alpha}-{\bf r}^{\rm CM}) \times {\bf
f}^{\alpha}=0,\label{eq:torque-free}
\end{equation}
where the center of mass (CM) position is defined as ${\bf r}^{\rm
CM}=\frac{1}{N} \sum_\alpha {\bf r}^{\alpha}$. We note that this
condition might not in general be satisfied, as in the case of a
recent experiment on magnetic doublets \cite{Pietro}. When it does
hold, however, it will introduce additional constraints on the type
of motion and conformations that we can prescribe for the system.
Finally, in sufficiently symmetric systems the torque-free
constraint might automatically be satisfied \cite{ag-pre}.

We now assume that the relative positioning of the body components
are prescribed, in a reference frame that moves with the average
position and orientation of the body. This reference frame, which we
call the ``body frame'' hereon, will be constant during one cycle of
the deformation in the body. As a result of the deformation, over
the period of one cycle the object is expected to be displaced by a
small amount due to a non-vanishing translational swimming velocity
and rotated slightly if there is a non-vanishing rotational velocity
as well. The combination of the displacement and rotation will
determine the new position and orientation of the body frame, which
will be used in the calculation of the next step of the motion and
so on. Therefore, in this picture the motions are grouped into
separate slow and fast degrees of freedom, in the sense that what is
happening over one deformation is cycle (fast degrees of freedom)
will be averaged to determine a net change in the slow degrees of
freedom that determine the overall average translation and rotation
of the swimmer through the liquid along its trajectory.

We now assume that the relative positioning of the body components
${R}^{\alpha \beta}_{i} \equiv r^{\alpha}_{i}-r^{\beta}_{i}$ are
known in the body frame, which means that the relative velocities
$v^{\alpha}_{i}-v^{\beta}_{i}={\dot R}^{\alpha \beta}_{i}$ are also
known.\footnote{Note that to get the actual form of ${R}^{\alpha
\beta}_{i}(t)$ from the internal motion of the object may require
calculation that involve the force-free and the torque-free
relations.} These relative positions and relative velocities need to
be prescribed in a such a way that all the necessary geometrical
constraints are satisfied, as for example, one cannot expect to have
arbitrary distances between a number of points that form a body of a
given shape.

If the shape of the object and the conformational changes are
sufficiently symmetric so that the object swims on average in a
rectilinear fashion, averaging the velocity of any tagged component
$\alpha$ over a complete cycle yields the total average
translational velocity of the body
\begin{equation}
\left \langle {\bf v}^{\alpha} \right \rangle={\bf V}^{\rm
trans},\label{eq:<v-1-i>}
\end{equation}
as the difference between the velocity of the $\alpha$ component and
that of the whole body will be in the form of relative deformations
that average out to zero. For a more general case the object will
have a rotational component superimposed with the translational one,
and the average velocity of the tagged body component in the body
frame will have the following form
\begin{equation}
\left \langle {\bf v}^{\alpha} \right \rangle={\bf V}^{\rm
trans}+{\mathbf \Omega}^{\rm rot} \times \left \langle \left({\bf
r}^{\alpha}-{\bf r}^{\rm CM} \right)\right
\rangle,\label{eq:<v-1-i>rot}
\end{equation}
where ${\mathbf \Omega}^{\rm rot}$ is the angular velocity vector of
the body about the center of mass. We can extract the translational
velocity as
\begin{equation}
{V}^{\rm trans}_{i}=\frac{1}{N} \sum_\alpha \left \langle
{v}^{\alpha}_{i} \right \rangle,\label{eq:Vtrans}
\end{equation}
and the rotational component of the velocity as
\begin{equation}
{\Omega}^{\rm rot}_{i}=I^{-1}_{ij} \sum_{\alpha} \epsilon_{jkl}
\left \langle \left({r}^{\alpha}_{k}-{r}^{\rm CM}_{k}\right)\right
\rangle\left \langle {v}^{\alpha}_{l} \right
\rangle,\label{eq:Omegarot}
\end{equation}
where
\begin{equation}
I_{ij}=\sum_{\alpha} \delta_{ij}\left \langle
\left({r}^{\alpha}_{k}-{r}^{\rm CM}_{k}\right)\right \rangle\left
\langle \left({r}^{\alpha}_{k}-{r}^{\rm CM}_{k}\right)\right
\rangle-\left \langle \left({r}^{\alpha}_{i}-{r}^{\rm
CM}_{i}\right)\right \rangle\left \langle
\left({r}^{\alpha}_{j}-{r}^{\rm CM}_{j}\right)\right
\rangle,\label{eq:Omegarot}
\end{equation}
is the average moment of inertia tensor for the object.

We can single out the velocity of one of the components, say
$\alpha=1$, and describe all of the velocities in terms of this and
the prescribed relative velocities, namely
$v^{\alpha}_{i}=v^{1}_{i}+{\dot R}^{\alpha 1}_{i}$. Putting this
back in equation (\ref{eq:oseen-2}), we find
\begin{equation}
v^{1}_{i}=-L^{-1}_{ij} \sum_{\alpha,\beta} N^{\alpha \beta}_{jk}
{\dot R}^{\beta 1}_{k},\label{eq:v-1-i}
\end{equation}
where $L_{ij}=\sum_{\alpha,\beta} N^{\;\alpha \beta}_{ij}$. Note
that one can also choose to specify the forces/tensions in the links
instead of the relative velocities. In this case it will be
straightforward to modify the formulation and calculate the
velocities. A more general framework would encompass prescriptions
relating stresses and deformations.

We can write the relative positioning of the components in the body
frame as ${R}^{\alpha \beta}_{i}(t)={R}^{\alpha \beta}_{0
\;i}+{u}^{\alpha \beta}_{i}(t)$, where ${u}^{\alpha \beta}_{i}$
denote the deformations of the body about the average shape
described by ${R}^{\alpha \beta}_{0 \;i}$. If we assume that the
deformations of the body are relatively small, we can expand
equation (\ref{eq:v-1-i}) in powers of the deformations and obtain
an expression for the instantaneous velocity of the tagged
($\alpha=1$) component of the body as
\begin{equation}
v^{1}_{i}(t)=\sum_{\alpha,\beta} A^{(1)\alpha \beta}_{ij} {\dot
u}^{\alpha \beta}_{j}+\sum_{\alpha,\beta,\gamma,\delta} B^{(1)\alpha
\beta \gamma \delta}_{ijk} {\dot u}^{\alpha \beta}_{j} {u}^{\gamma
\delta}_{k}+\sum_{\alpha,\beta,\gamma,\delta,\mu,\nu} C^{(1)\alpha
\beta \gamma \delta \mu \nu}_{ijkl} {\dot u}^{\alpha \beta}_{j}
{u}^{\gamma \delta}_{k} {u}^{\mu \nu}_{l}+\cdots,\label{eq:v-i}
\end{equation}
where the coefficients $A^{(1)\alpha \beta}_{ij}$, $B^{(1)\alpha
\beta \gamma \delta}_{ijk}$, $C^{(1)\alpha \beta \gamma \delta \mu
\nu}_{ijkl}$, {\em etc.} are purely geometrical pre-factors ({\em
i.e.} involving only the characteristic length scales describing the
shape of the body). Averaging over a full cycle, the contribution
due to the linear terms ${\dot u}^{\alpha \beta}_{j}$ and the
symmetric combinations ${\dot u}^{\alpha \beta}_{j} {u}^{\gamma
\delta}_{k}+{\dot u}^{\gamma \delta}_{k} {u}^{\alpha \beta}_{j} =d
({u}^{\alpha \beta}_{j} {u}^{\gamma \delta}_{k})/dt$ vanish.
Therefore, to the leading order, we find the average swimming
velocity as
\begin{equation}
{V}^{\rm trans}_{i}=\frac{1}{N} \sum_\mu \left \langle {v}^{\mu}_{i}
\right \rangle=\frac{1}{2}\sum_{\alpha,\beta,\gamma,\delta}
B^{\alpha \beta \gamma \delta}_{ijk} \left \langle{\dot u}^{\alpha
\beta}_{j} {u}^{\gamma \delta}_{k}-{\dot u}^{\gamma \delta}_{k}
{u}^{\alpha \beta}_{j} \right
\rangle=\sum_{\alpha,\beta,\gamma,\delta} B^{\alpha \beta \gamma
\delta}_{ijk} \left \langle \frac{\Delta {\cal A}^{\alpha \beta
\gamma \delta}_{jk}}{\Delta t} \right \rangle,\label{eq:<v>-i}
\end{equation}
where $\Delta {\cal A}^{\alpha \beta \gamma \delta}_{jk}$ is the
area element enveloped by the trajectory of the system in the
$(u^{\alpha \beta}_{j},u^{\gamma \delta}_{k})$ space, and $B^{\alpha
\beta \gamma \delta}_{ijk}=\frac{1}{N} \sum_\mu B^{(\mu)\alpha \beta
\gamma \delta}_{ijk}$. Note that ${\Delta {\cal A}^{\alpha \beta
\gamma \delta}_{jk}}/{\Delta t}$ is not a complete time derivative,
and its average over a a full cycle does not vanish. A similar
expression can be written for the angular velocity. The averaging
here denotes time averaging if the conformation of the system is
prescribed. If, however, the system undergoes stochastic
conformational changes, the averaging needs to be performed over the
distribution of the various conformations. The formulation needed to
carry out this step of the calculation is developed in the next
section.

\section{Kinetics in the Conformation Space} \label{sec:kinetics}

\begin{figure}[t]
\includegraphics[width=16pc]{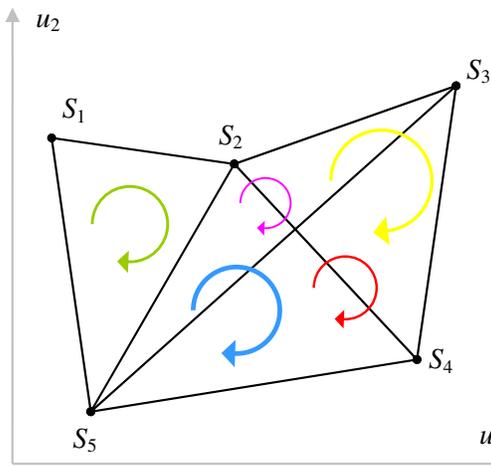}\hspace{0pc}%
\begin{minipage}[b]{19pc}\caption{A typical conformation subspace
describing the shape of the swimmer. Distinct conformational states
are identified and connected to one another when transitions are
permissible, making a graph. The swimming velocity will be
determined by the sum of the currents in each loop of the graph
(denoted by different colors here), weighted by the area of each
loop, correspondingly [see equation (\ref{eq:dAdt})].}
\label{fig:gen}
\end{minipage}
\end{figure}

Let us now consider a conformation subspace of the system
corresponding to two representative deformations $u_1$ and $u_2$
(see figure \ref{fig:gen}). Since we aim to model molecular systems,
we should take into account the stochastic nature of the
conformational changes and not prescribe a deterministic trajectory
for the deformation of the system. We identify distinct
conformational states of the system, denoted as $S_n$, and construct
a kinetic description where the deformations of the system are
described by transitions between these states with given rates,
assuming that they occur one at a time and do not overlap with each
other. We denote the probability of finding the system in $S_n$ as
$P_n$ and the rate for transition $m \rightarrow n$ as $k_{nm}$.
These probabilities are normalized as $\sum_n P_n=1$. Connecting the
states that have permissible transitions between them with links, we
find a graph that characterizes the conformational kinetics of the
system in each subspace, as seen in figure \ref{fig:gen}. To every
link, we can attribute a probability current
\begin{equation}
J_{<nm>}=k_{mn}P_n-k_{nm}P_m,
\end{equation}
and at stationary state we can impose the continuity of current at
every node, namely
\begin{equation}
\sum_m J_{<nm>}=0.
\end{equation}
Solving the system of equations, we can find all probabilities and
currents, and in particular the currents $J(\alpha)$ running through
all the loops in the graph (see figure \ref{fig:gen}). We can then
write
\begin{equation}
\left \langle \frac{\Delta {\cal A}}{\Delta t} \right
\rangle=\sum_{\alpha} {\cal A}(\alpha) J(\alpha),\label{eq:dAdt}
\end{equation}
where ${\cal A}(\alpha)$ is the area enclosed by loop $\alpha$ in
the conformation subspace. Equation (\ref{eq:dAdt}) shows that the
contributions from the different loops act together analogously to
circuits {\em in parallel}, and therefore, it will be the fastest
route that will determine the effective swimming velocity. In each
loop, however, the different legs are connected {\em in series}, and
the slowest kinetic leg will control the contribution to the
effective swimming velocity from each loop (see the example below).

\section{Example: Three-Sphere Swimmer Model} \label{sec:3SS}

\begin{figure}[t]
\includegraphics[width=18pc]{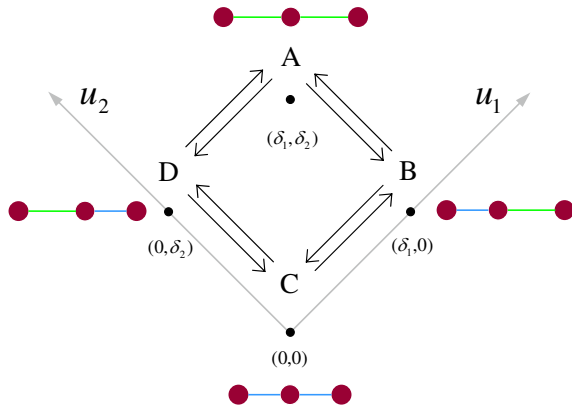}\hspace{0pc}%
\begin{minipage}[b]{18pc}\caption{Conformation space of the three-sphere
swimmer model. This minimal model involves only one loop. The
convention is such that a net swimming to the right requires the
system to make more cycles in the clockwise direction than in the
counterclockwise direction.} \label{fig:config}
\end{minipage}
\end{figure}

We now focus on the specific example of a three-sphere swimmer model
\cite{3SS}. We define the conformation space of the swimmer using
the two variables $(u_1,u_2)$ that describe the longitudinal
deformation of the two arms of the swimmer. We assume that the two
arms can be in the two states corresponding to either $u_i=0$ or
$u_i=\delta_i$, and transit from one to the other in an almost
instantaneous fashion. This means that the configuration space of
the swimmer will be made of four distinct states as shown in figure
\ref{fig:config}, defined by different values of the pair
$(u_1,u_2)$, namely: state A for $(\delta_1,\delta_2)$, state B for
$(\delta_1,0)$, state C for $(0,0)$, and state D for $(0,\delta_2)$.
We then assign transition rates to the system, corresponding to the
average rate of opening and closing of the arms along the cycle
\begin{equation}
{\rm A}\xrightleftharpoons[k_{AB}]{k_{BA}} {\rm
B}\xrightleftharpoons[k_{BC}]{k_{CB}}{\rm
C}\xrightleftharpoons[k_{CD}]{k_{DC}}{\rm
D}\xrightleftharpoons[k_{DA}]{k_{AD}}{\rm A}. \label{eq:cycle}
\end{equation}
Note that in this simple example there is only one loop in the
conformation space graph of the system (figure \ref{fig:config}).

We can now calculate the swimming velocity as a function the
transition rates. Using the general formulation described in
Sections \ref{sec:hydro} and \ref{sec:kinetics}, we find
\begin{equation}
V= K\delta_1\delta_2 J.\label{eq:V-def2}
\end{equation}
where
\begin{math} K=\frac{a}{3}
\left[\frac{1}{\ell_1^2}+\frac{1}{\ell_2^2}-\frac{1}{(\ell_1+\ell_2)^2}\right]
\end{math}
with $\ell_1$ and $\ell_2$ being the undeformed lengths of the two
arms and $a$ being the radius of the spheres \cite{ag-pre}. The
probability current $J$ is a function of the transition rates, which
can be obtained from the following straightforward algebra. At
steady state, the current conservation equations can be written as
$J=k_{BA}P_A-k_{AB}P_B=k_{CB}P_B-k_{BC}P_C=k_{DC}P_C-k_{CD}P_D=k_{AD}P_D-k_{DA}P_A$,
which provide us with four equations for the current and the four
probabilities, which are also normalized as $P_A+P_B+P_C+P_D=1$.
Solving the system of linear equations, we find
\begin{equation}
J=\frac {k_{AD}k_{DC}k_{CB}k_{BA}-k_{AB}k_{BC}k_{CD}k_{DA} } {\sum
_{\rm replace \;A \;by \;B,\;C,\;D} (k_{AD}k_{DC}k_{CB}
+k_{AB}k_{BC}k_{CD} +
k_{AB}k_{AD}k_{DC}+k_{AD}k_{AB}k_{BC})}.\label{eq:J}
\end{equation}
Equations (\ref{eq:V-def2}) and (\ref{eq:J}) give the swimming
velocity of the three-sphere swimmer \cite{ag}.

From equation (\ref{eq:J}) it is clear that if detailed balance
holds, then $J$ vanishes as the numerator is zero. Using the average
steady state current, we can deduce the average period of completing
one full cycle of the motion along the
A$\rightarrow$B$\rightarrow$C$\rightarrow$D$\rightarrow$A loop as
\begin{equation}
T=J^{-1}.
\end{equation}
We can gain a useful insight by looking at the particular limit
where the forward rates are all much higher than the corresponding
backward ones ($k_{BA} \gg k_{AB}$, {\em etc.}). In this limit, we
find
\begin{equation}
T=k_{AD}^{-1}+k_{DC}^{-1}+k_{CB}^{-1}+k_{BA}^{-1},
\end{equation}
which means that the period for a full cycle is the sum of the time
intervals needed to complete each leg of the cycle.

As another example, we can assume that all of the equilibrium
$k_{\beta \alpha}$'s are equal to $1$ (for simplicity), and that by
external action only one of them is modified as $k_{BA}=1+\epsilon$.
In this case, one can show that equation (\ref{eq:J}) yields
\begin{equation}
J=\frac{\epsilon}{16+6\epsilon},
\end{equation}
which leads to a velocity proportional to the perturbation for small
values of $\epsilon$ and independent of it if the perturbation is
very large. The linear dependence can be easily understood for a
system that is only slightly driven out of equilibrium, and the
saturation at large perturbations is because the cycling will then
be limited by the other three unperturbed transitions. In general,
one can see that the slowest leg of the reaction controls the
average rate of full cyclic motion, which suggests the
interpretation that in each loop the different legs are connected in
series, in analogy to circuits.

\section{Conclusion} \label{sec:conc}

We have presented a general formulation that can be used in studying
the swimming of a small object that undergoes stochastic
deformations. The program to follow to this end has two stages: (1)
treat the deformations as prescribed and follow the hydrodynamic
formulation of Section \ref{sec:hydro} to calculate the average
swimming velocity in terms of the relevant deformation variables.
(2) Construct the conformation space of the system based on the
deformation variables and follow the statistical mechanical
description of Section \ref{sec:kinetics} to work out the
contributions to the net swimming velocity by various {\em modes} of
swimming defined as loops in the conformation space. We found that a
useful circuits analogy can be invoked to describe the efficiency of
the swimming, with two notable features: (1) the different modes of
swimming can be effectively considered to act {\em in parallel},
which means that their contributions will be independently added to
each other to yield the net swimming velocity and therefore the {\em
fastest} route will be the dominant mode of swimming controlling the
velocity. (2) In each loop, the different kinetic legs could be
considered as acting {\em in series} with respect to one another,
which means that the {\em slowest} kinetic leg will control the net
contribution to the velocity by the loop.

The formulation also allows us to study the effect of an external
force or load on the performance of swimmers. External forces both
add to the hydrodynamic drag and also affect the performance of the
swimming strokes as activated moves, as the deformations will
involve doing work against or being helped by forces endured by the
arms. These forces will modify the transition rates, and their
effects can be readily accommodated by using the force-dependent
rates in the kinetic formulation. This effect has been studied for
the three-sphere swimmer model, which has revealed that the
performance of the motor strongly depends on where the force is
exerted \cite{ag}. This shows that for such small swimmers, the
concept of a generic force--velocity response breaks down, which
might have interesting implications for designing molecular
swimmers.


\section*{References}
\begin{thebibliography}{10}

\bibitem{howard}
J. Howard, {\em Mechanics of Motor Proteins and the Cytoskeleton}
(Sinauer, New York, 2000)

\bibitem{leigh-etal}
E.R. Kay, D.A. Leigh, and F. Zerbetto, Angew. Chem. Int. Ed. {\bf
46}, 72 (2007)

\bibitem{taylor}
G.I. Taylor, Proc. Roy. Soc. London {\bf A 209}, 447-461 (1951)

\bibitem{purcell1}
E.M. Purcell, American Journal of Physics {\bf 45}, 3-11 (1977)

\bibitem{3SS}
A. Najafi and R. Golestanian, Phys. Rev. E {\bf 69}, 062901 (2004)

\bibitem{josi}
J.E. Avron {\em et al.}, Phys. Rev. Lett. {\bf 93}, 186001 (2004)

\bibitem{drey}
R. Dreyfus {\em et al.}, Eur. Phys. J. B {\bf 47}, 161(2005)

\bibitem{igor}
I.M. Kulic {\em et al.} Europhys. Lett. {\bf 72}, 527 (2005)

\bibitem{lee}
A. Lee {\em et al.}, Phys. Rev. Lett. {\bf 95}, 138101 (2005)

\bibitem{ali2}
A. Najafi and R. Golestanian, J. Phys.: Condens. Matter {\bf 17},
S1203 (2005)

\bibitem{feld}
B.U. Felderhof, Phys. Fluids {\bf 18}, 063101 (2006)

\bibitem{holger}
E. Gauger and H. Stark, Phys. Rev. E {\bf 74}, 021907 (2006)

\bibitem{lesh}
A.M. Leshansky, Phys. Rev. E {\bf 74}, 012901 (2006)

\bibitem{peko}
D. Tam and A.E. Hosoi, Phys. Rev. Lett. {\bf 98} 068105 (2007)

\bibitem{yeomans1}
D.J. Earl {\em et al.}, J. Chem. Phys. {\bf 126} 064703 (2007)

\bibitem{anna}
C.M. Pooley and A.C. Balazs, Phys. Rev. E {\bf 76}, 016308 (2007)

\bibitem{yeomans2}
C.M. Pooley {\em et al.}, Phys. Rev. Lett. {\bf 99}, 228103 (2007)

\bibitem{yeomans3}
G.P. Alexander and J.M. Yeomans Europhys. Lett. {\bf 83}, 34006
(2008)

\bibitem{yeomans4}
G.P. Alexander {\em et al.}, Phys. Rev. E {\bf 78}, 045302 (R)
(2008)


\bibitem{lauga2}
E. Lauga and D. Bartolo, Phys. Rev. E {\bf 78} 030901 (R) (2008)

\bibitem{ag-pre}
R. Golestanian and A. Ajdari, Phys. Rev. E {\bf 77}, 036308 (2008)

\bibitem{epje}
R. Golestanian, Eur. Phys. J. E {\bf 25}, 1 (R) (2008)

\bibitem{ag}
R. Golestanian and A. Ajdari, Phys. Rev. Lett. {\bf 100}, 038101
(2008)

\bibitem{Dreyfus}
R. Dreyfus {\em et al.}, Nature {\bf 437}, 862 (2005)

\bibitem{Pietro}
P. Tierno {\em et al.}, Phys. Rev. Lett. {\bf 101}, 218304 (2008)

\bibitem{paxton}
W.F. Paxton {\em et al.}, J. Am. Chem. Soc. {\bf 126}, 13424 (2004)

\bibitem{ozin}
S. Fournier-Bidoz {\em et al.}, Chem. Comm., 441-443 (2005)

\bibitem{mano}
N. Mano and A. Heller, J. Am. Chem. Soc. {\bf 127}, 11574 (2005)

\bibitem{jon}
J.R. Howse {\em et al.}, Phys. Rev. Lett. {\bf 99}, 048102 (2007)

\bibitem{gla-1}
R. Golestanian {\em et al.}, Phys. Rev. Lett. {\bf 94}, 220801
(2005)

\bibitem{gla-2}
R. Golestanian {\em et al.}, New J. Phys. {\bf 9}, 126 (2007)

\bibitem{kapral}
G. R\"uckner and R. Kapral, Phys. Rev. Lett. {\bf 98}, 150603 (2007)

\bibitem{Oseen}
C.W. Oseen, {\em Neuere Methoden und Ergebnisse in der Hydrodynamik}
(Akademishe Verlagsgesellschaft, Leipzig, 1927).

\end{thebibliography}
\end{document}